\documentclass[aps,prl,twocolumn,twoside,floatfix,superscriptaddress,preprintnumbers]{revtex4}
\usepackage{graphicx}
\newcommand{\bea}{\begin{eqnarray}}
\newcommand{\beq}{\begin{equation}}
\newcommand{\eea}{\end{eqnarray}}
\newcommand{\eeq}{\end{equation}}

\newcommand{\eps}{ \epsilon }

\newcommand{\epsY}{ {\epsilon_Y} }

\newcommand{\epsL}{ {\epsilon_L} }
\newcommand{\epsR}{ {\epsilon_R} }
\newcommand{\epsLR}{ {\epsilon_{LR}} }
\newcommand{\tgb}{ t_\beta }
\newcommand{\ttgb}{ t_\beta^2 }

\newcommand{\cotb}{ t_\beta^{-1} }

\newcommand{\Vb}{ \overline{V} }
\newcommand{\yb}{ \overline{y} }
\newcommand{\mb}{ \overline{m} }

\newcommand{\uL}{ {u_L} }

\newcommand{\dL}{ {d_L} }
\newcommand{\dR}{ {d_R} }

\newcommand{\uRb}{ {{\overline u}_R} }

\newcommand{\dRb}{ {{\overline d}_R} }

\newcommand{\tHdLmb}{  { \overline{\tilde{H}} \!\!\ _{1L}^-} }

\newcommand{\tHuRmb}{  { \overline{\tilde{H}} \!\!\ _{2R}^-} }

\newcommand{\lsim}{\stackrel{<}{_\sim}}

\newcommand{\epsYt}{ \epsilon_Y }

\begin{document}

\title{``Flavored'' Electric Dipole Moments in Supersymmetric Theories}

\author{J.~Hisano}
\affiliation{ICRR, University of Tokyo, Kashiwa 277-8582, Japan}
\author{M.~Nagai}
\affiliation{ICRR, University of Tokyo, Kashiwa 277-8582, Japan}
\author{P.~Paradisi}
\affiliation{Departament de F\'{\i}sica Te\`orica and IFIC, Universitat de
Val\`encia-CSIC, E-46100, Burjassot, Spain}

\begin{abstract}
  The Standard Model (SM) predictions for the hadronic electric dipole
  moments (EDMs) are well far from the present experimental
  resolutions, thus, the EDMs represent very clean probes of New
  Physics (NP) effects. Especially, within an MSSM framework with
  flavor-changing (but not necessarily CP violating) soft terms, large
  and potentially visible effects to the EDMs are typically expected.
  In this Letter we point out that, beyond-leading-order (BLO)
  effects, so far neglected in the literature, dominate over the
  leading-order (LO) effects in large regions of the parameter space,
  hence, their inclusion in the evaluation of the hadronic EDMs is
  unavoidable.
\end{abstract}

\maketitle

The Standard Model (SM) of elementary particles has been very
successfully tested at the loop level both in (flavor-conserving)
electroweak (EW) physics at the LEP and also in low-energy flavor
physics. In particular, the two $B$ factories have allowed an accurate
determination of all the relevant parameters describing quark-flavor
mixing within the SM. In this way, the overall picture of particle
physics is a bit frustrating as far as the search for physics beyond
the SM is concerned since little room is left for NP effects.

On the other hand, it is a common belief that the SM has to be
regarded as an effective field theory, valid up to some still
undetermined cut-off scale $\Lambda$ above the EW scale. Theoretical
arguments based on a natural solution of the hierarchy problem suggest
that $\Lambda$ should not exceed a few TeV, an energy scale that will
be explored at the upcoming LHC.

Besides the direct search for NP at the scale (the so-called
{\em high-energy frontier}), a complementary and equally important
tool to shed light on NP is provided by high-precision low-energy
experiments (the so-called {\em high-intensity frontier}). The latter
are suitable in determining the symmetry properties of the underlying
NP theory. Unfortunately, the hadronic uncertainties and the overall
good agreement of flavor-changing neutral current (FCNC) data with the
SM predictions prevent any conclusive evidence of NP effects in the
quark sector. In this respect, the FCNC phenomenology in the lepton
sector is definitively more promising.  In fact, the extreme
suppression predicted by the SM (with massive neutrinos) for processes
like $\ell_i\to\ell_j\gamma$ implies that any experimental evidence
for $\ell_i\to\ell_j\gamma$ would unambiguously point towards a NP
signal.

The hadronic EDMs also offer a unique possibility to shed light in NP,
given their strong suppression within the SM and their high
sensitivities to NP effects. The minimal supersymmetric SM (MSSM),
that is probably the most motivated model beyond the SM, exhibits
plenty of CP-violating phases \cite{pospelov} able to generate the
hadronic EDMs at an experimentally visible level \cite{expedm}. These
new CP phases may be introduced both in the flavor-conserving and in
the flavor-changing soft SUSY breaking terms. In the latter case, the
``flavored'' EDMs are strongly constrained and/or correlated with FCNC
processes from $K$ and $B$ physics.

In this Letter, we analyze the predictions for the ``flavored'' quark EDMs
and Chromo-EDMs (CEDM), which contribute to the hadronic EDMs, at the BLO.

At the LO, the hadronic EDMs are generated by the one-loop exchange
of gluinos ($\tilde{g}$) and charginos ($\tilde{\chi}^\pm$) with squarks.
The dominant BLO contributions are computed by including all the one-loop
induced ($\tan\beta$-enhanced) non-holomorphic corrections for the charged
Higgs ($H^\pm$) couplings with fermions and for the $\tilde{\chi}^\pm$
/$\tilde{g}$ couplings with fermions-sfermions.
The above effective couplings lead to the generation of $H^\pm$ effects to
the (C)EDMs, absent at the LO, via the one-loop $H^\pm$/top-quark exchange.
Moreover, the chargino contributions, suppressed at the LO by the light quark
masses, are strongly enhanced at the BLO by the heaviest-quark Yukawa
couplings. Finally, also the gluino effects receive large BLO contributions that
are comparable, in many cases, to the LO ones.
As a result, BLO effects, so far neglected in the literature, dominate over the
LO effects in large regions of the SUSY parameters space.
Hence, their inclusion in the evaluation of the flavored (C)EDMs is mandatory.

The SM sources of CP violation are the QCD theta term
$\overline{\theta}$ and the unique physical phase contained in the CKM
matrix, $V$.  However, a Peccei-Quinn symmetry \cite{peccei} is commonly
assumed making $\overline{\theta}$ dynamically suppressed.
In this way, the hadronic EDMs can be generated by the only CP-violating
phase of the CKM. The best way to describe them is provided by the
Jarlskog invariants (JIs) that are a basis-independent measure of CP
violation \cite{Jarlskog}.  In the SM, the $i$-th quark EDMs $d_{q_i}$
and CEDMs $d_{q_i}^c$ are induced by the flavor-conserving JIs,
\beq
J^{(d_i,u_i)}_{\rm SM}= {\rm Im} \left\{Y_d[Y_d, Y_u]Y_u y_{d,u}\right\}_{ii}\,,
\label{inv_SM}
\eeq
where $y_d$($y_u$) is the down(up)-type quark Yukawa coupling constant and 
$Y_{d(u)} \equiv y_{d(u)}y_{d(u)}^\dagger$. Since $J^{(d_i,u_i)}_{\rm SM}$ 
are of the ninth order in the Yukawa coupling constants, the quark (C)EDMs 
are highly suppressed at the level of $\sim 10^{-(33-34)}~ e\,{\rm cm}$.

So, we conclude that the SM expectations for the hadronic EDMs are well
below the actual and expected future experimental resolutions \cite{expedm}.

Within a SUSY framework, CP-violating sources may naturally appear
after the SUSY breaking through (i) flavor-conserving $F$-terms (such
as the $B$ parameter in the Higgs potential or the $A$ terms for
trilinear scalar couplings) and (ii) flavor-violating $D$-terms (such
as the squark and slepton mass terms).  In the case (i), the
experimental bounds on the EDMs constrain the phases $\phi_{A,B}$ to
be very close to zero: this naturalness problem is known as the SUSY
CP problem \cite{pospelov}.  In this respect, a mechanism leading to a
natural suppression of the (C)EDMs would be desirable and, indeed, this
is what happens in the case (ii).

The presence of flavor structures in the soft sector, that we
parameterize as usual by means of the mass insertion (MI) parameters
\cite{GGMS}
$(\delta_{AA}^q)_{ij}\equiv(m^2_{\tilde{q}_{AA}})_{ij}/{m}_{\tilde{q}}^2$
($A=L/R$), generally leads to FCNC transitions and to
(flavor-conserving/violating) CP-violating phenomena. As a natural
consequence, the hadronic EDMs are generated and they turn out to be
intimately linked to FCNC processes, as they both arise from the same
source.

The size and the pattern of the MIs are unknown, unless we assume
specific models.  They are regulated by the SUSY breaking mechanism
and by the interactions of the high-energy theories beyond the MSSM.
In this way, it makes sense to study the individual impact of
different kinds of MIs on the low-energy observables.
When only $(\delta_{LL}^q)_{ij}\ne 0$, the following flavor-conserving JI,
which contributes to the down-quark (C)EDMs, shows up,
\begin{eqnarray}
J^{(d_i)}_{LL}= {\rm Im} \left\{[Y_u,\delta_{LL}^q]y_d\right\}_{ii}\,,
\label{inv_LL}
\end{eqnarray}
while, if $(\delta_{RR}^d)_{ij}\ne 0$, we generate the invariant
\begin{eqnarray}
J^{(d_i)}_{RR}= {\rm Im} \left\{Y_u y_d\delta_{RR}^d \right\}_{ii}\,.
\label{inv_RR}
\end{eqnarray}
Both $J^{(d_i)}_{LL}$ and $J^{(d_i)}_{RR}$ are of the third order in
the Yukawa coupling constants. If both $(\delta_{RR}^d)_{ij}\ne 0$ and
$(\delta_{LL}^d)_{ij}\ne 0$, the (C)EDMs emerge at the one-loop level
through
\begin{eqnarray}
J^{(d_i)}_{LR}= {\rm Im} \left\{\delta_{LL}^q y_d\delta_{RR}^d
\right\}_{ii}.
\label{inv_LR}
\end{eqnarray}
The invariant $J^{(d_i)}_{LR}$ is proportional to only one Yukawa
coupling constant that is relative to the heaviest quark generation.
In the minimal flavor violation hypothesis \cite{MFV0,MFV}, where the
MIs are given by the SM Yukawa couplings $y_u$ and $y_d$, all the
above JIs are suppressed at the same level of $J^{(d_i)}_{\rm SM}$ in
the SM.  So, large effects to the (C)EDMs can be generated only if new
flavor structures in addition to the CKM are present. However, even
if these new flavor structures do not introduce any new source of CP
violation, $J^{(d_i)}_{LL}$ and $J^{(d_i)}_{RR}$ are generally
non-vanishing thanks to their dependence on the CKM phase.

The JI contributing to the up-quark (C)EDMs are simply obtained from
the corresponding down-quark JI by exchanging the suffixes $u$ and
$d$.  For simplicity, in the following discussion we focus our
attention on the down-quark (C)EDMs; the extension to the up-quark
case is straightforward.

The effective Lagrangian necessary to evaluate all the relevant BLO effects
to the (C)EDMs includes effective couplings of $H^\pm$ with fermions and
of $\tilde{\chi}^\pm$ ($\tilde{g}$) with fermion-sfermion.

The determination of the $\tan\beta$-enhanced effects passes through the
following steps \cite{MFV,EFFL}: {\it i)} evaluation of the effective
dimension-four operators appearing at the one-loop level which modify
the Yukawa couplings; {\it ii)} expansion of the off-diagonal squark
mass terms by means of the MI approximation; {\it iii)}
diagonalization of the quark mass terms and derivation of the relevant
effective interactions.

Starting from step {\it i)}, the interaction Lagrangian for the Higgs and
fermion fields, in the SU(2)$\times$U(1) symmetric limit, is expressed by
\begin{eqnarray} \label{lag_NH}
 {\cal L} =
  \uRb_i
  \left[
   {\yb_u}_i \Vb_{ij} H_2
   -
   (\eps^u\Vb)_{ij} H^\dagger_1
  \right]
  {q_L}_j 
\nonumber    \\ 
+  \dRb_i
  \left[
   {\yb_d}_i \delta_{ij} H_1
   -
  \eps^d_{ij} H^\dagger_2
  \right]
  {q_L}_j
  + {\rm h.c.}\,,
\end{eqnarray}
where $\eps^q_{ij}$s are the non-holomorphic radiative corrections
appearing when heavy SUSY particles are integrated out from the
effective theory \cite{MFV,EFFL,HRS}, while $\Vb$ ($\yb_f$) are the
CKM matrix (Yukawa couplings) defined in the ``bare'' CKM basis,
namely the CKM basis as defined before the inclusion of the
$\eps^q_{ij}$ corrections.

Passing to step {\it ii)} we derive, after the electroweak symmetry
breaking, the radiative correction to the down-quark mass matrix
$(\delta m_d)_{ij}$ as follows
\begin{eqnarray} \label{mass_correction}
 (\delta m_d)_{ij}
 \!\!\!\! ~~\simeq~ \!\!\!\!
  \Biggl[
   \mb_{d_i}\eps\delta_{ij}
   +
   \overline{m}_{d_i}\!
   \left(
   \epsY(\Vb^\dagger\!\Delta\Vb)_{ij}
   -\eps_L (\delta^d_{LL})_{ij}
   \right) 
 \nonumber\\
% &&
   -
  \eps_R (\delta^d_{RR})_{ij}\mb_{d_j}
   +
  \eps_{LR} \overline{m}_b
  (\delta^d_{RR}\Delta\delta^d_{LL})_{ij}
  \Biggr]
  \tgb\,.
\label{quarkmasses}
\end{eqnarray}
In the $(\delta m_d)_{ij}$ evaluation, we included the effects from
flavor-violation sources (MIs) in the squark mass matrices. In
Eq.~(\ref{quarkmasses}) $\tgb \!\!=\!\!\tan\beta$,
$\Delta\!\!=\!\!{\rm diag}(0,~0,~1)$ and, for equal SUSY masses and
$\mu>0$, it turns out that
$6\eps_{LR}\!=\!3\eps_{L}\!=\!3\eps_{R}\!=\!\eps\!=\!\alpha_s/3\pi$
and $\eps_Y=-A_t/|A_t|\!\times\!(y_t^2/32\pi^2)$ with $A_t$ defined
in the convention where the left-right stop mass term is
$-m_t(A_t+\mu/\tgb)$; moreover, in Eq.~(\ref{quarkmasses}), as in the
rest of this Letter, hat and bar symbols refer to diagonal matrices
and bare quantities, respectively.

Passing to step {\it iii)} we define the ``physical'' CKM basis through
the unitary transformations
\begin{equation}
d'_L = V_{d_L}d_{L}, ~~ u'_L = V_{u_L} \Vb~u_{L},
~~ d'_R = e^{-i\hat{\theta}_d} V_{d_R}d_{R}\,,
\label{rotations}
\end{equation}
so that the ``physical'' CKM matrix is given by $V=V_{u_L}\Vb V_{d_L}^\dagger$. Allowing
for complex entries in $(\delta m_d)_{ij}$, the phase rotations $\exp({-i\hat{\theta}_d})$
are introduced to make the quark masses real.

Expanding $V_{d_{R,L}}$ at the first order around the diagonal, {\it i.e.},
$V_{d_{R,L}}\simeq {\bf 1} + \delta V_{d_{R,L}}$, we find, by means of
Eq.~(\ref{mass_correction}), the following expressions,
\begin{eqnarray} \label{Unitary_components}
 (\delta V_{d_L})_{3i}
 \!\! &\simeq& \!\!
  - \frac{\epsL\tgb}{1+\eps\tgb}(\delta_{LL}^d)_{3i}
  + \frac{\epsY\tgb}{1+\eps\tgb}V_{3i}\,,
 \nonumber\\
 (\delta V_{d_R})_{i3}
 \!\! &\simeq& \!\!
  \frac{\epsR\tgb}{1+\eps\tgb}(\delta_{RR}^d)_{i3}\,.
\end{eqnarray}
The phase parameters ${\theta_d}_{1,2}$ in $\hat{\theta}_d$ are
derived from
\begin{eqnarray}\label{mass_matching}
 {m_d}_i e^{i{\theta_d}_i} \simeq {\mb_d}_i + (\delta m_d)_{ii}
  + m_b (\delta V_{d_R})_{i3} (\delta V_{d_L}^\star)_{i3}\,,
\end{eqnarray}
and ${\theta_d}_3\simeq0$.
$V_{u_L}$ is obtained by $V_{d_L}$ by exchanging $\Vb$ with $\Vb^\dagger$
and $\tan\beta$ with $\cotb$. Finally, we can derive the effective $H^\pm$
couplings with fermions
\begin{eqnarray}
 \bar{t}_L \dR_i H^+\!\!\!\!&\to&\!\!
  y_{d_i}
  \bigg[\frac{{\mb_d}_i}{{m_d}_i} e^{i{\theta_d}_i} \Vb_{3i}
  + \frac{{\mb_b}}{{m_d}_i} e^{i{\theta_d}_i} V^{*i3}_{d_R}\bigg]\,,
 \\
 \bar{t}_R \dL_i H^+ \!\!\!\!&\to&\!\!
  y_{t}\cotb
  \bigg[(1\!-\!\eps\tgb) V_{3i}-\ttgb\sum_{j\neq 3}V^{3j}_{\uL}V_{ji}\bigg]\,.
\label{higgscoupl}
\end{eqnarray}
The charged Higgsino ($\tilde{H}^\pm$) couplings are also given as
\begin{eqnarray}
 \tilde{t}_{R}^* \tHuRmb \dL_i
  &\to&
  - y_{t} (V_{u_L}^\dagger V)_{3i}\,,
  \\
 \tilde{t}_{L}^* \tHdLmb \dR_i
  &\to&
  (\overline{V} \hat{\yb}_d V_{d_R}^\dagger)_{3i}\, e^{-i\theta_{d3}}\,,
\label{charginocoupl}
\end{eqnarray}
where $i=1,2$. The $\tilde{g}$ interactions are described by
\begin{eqnarray}
{\tilde d}^{*}_{Li}\,\overline{{\tilde g}}_{R}^{a}\,d_{Lj}
&\to&
\sqrt{2} g_s(V^{\dagger}_{d_{L}})_{ij}\,,
\nonumber \\
{\tilde d}^{*}_{Ri}\,\overline{{\tilde g}}_{L}^{a}\,d_{Rj}
&\to&
-\sqrt{2} g_s (V^{\dagger}_{d_{R}})_{ij}\, e^{-i\theta_{di}}\,.
\label{gluinocoupl}
\end{eqnarray}
The above Feynman rules have been computed performing the rotations
for the quark fields (see Eq.~(\ref{rotations})) in the ``bare''
Lagrangian and implementing, at the same time, the vertex corrections
to the relevant interactions. Other couplings, such as those for the
wino or the bino, are also derived in a similar way.

Let us notice that the $H^\pm$ and $\tilde{H}^\pm$ couplings with the
right-handed down quarks can be proportional to the heaviest-quark
Yukawa coupling when the right-handed squark mixing is non-vanishing.
The last mechanism provides the main enhancement factor for the BLO
contributions. In addition, at the BLO, $\tilde{g}$ interactions develop
flavor- and/or CP-violating couplings that also lead to sizable
effects on the (C)EDMs.

Although our numerical results have been obtained including the full
set of contributions, in the following, for simplicity, we report only
the dominant contributions to the hadronic EDMs.
In particular: {\it  i)} we neglect the effects proportional to
$J^{(d_i)}_{LL}$ because suppressed by a factor of $m_{d_i}/m_b$ compared
to the dominant contributions;
{\it ii)} we neglect the sub-leading effects provided by the electroweak 
couplings $g_1$ and $g_2$.
%A comprehensive analysis of the full computation will appear in an upcoming communication.
For later convenience, let us define $\omega_{1,2}$ as
\begin{eqnarray}
\omega_1 &=& {\rm Im}\left[(\delta^d_{LL})_{i3}(\delta^d_{RR})_{3i}\right]\,,\\
\omega_2 &=& {\rm Im}\left[ V^*_{3i}(\delta^d_{RR})_{3i} \right]\,,
\end{eqnarray}
to which $J^{(d_i)}_{LR}$ and $J^{(d_i)}_{RR}$  are proportional, respectively.
The gluino/squarks contribution to the (C)EDMs is
\begin{eqnarray}
 \left\{ \frac{d_{d_i}}{e},~d^c_{d_i} \right\}_{\tilde g}
 \!\!\!\!\!~\simeq~\!\!
 \frac{\alpha_s}{4\pi}
 \frac{m_b}{m_{\tilde{q}}^2}
 \frac{m_{\tilde g} \mu}{m_{\tilde{q}}^2}
 \frac{\tgb}{1\!+\!\eps\tgb}
 \bigg[
 \omega_1\, f_{\tilde g}^{(3)}(x) +
 \nonumber \\
 \!\!\!
f_{\tilde g}^{(2)}\!(x)
 \!\bigg(\!\frac{(\eps_L+\eps_R)\tgb}{1\!+\!\eps\tgb}\omega_1
\!-\!
\frac{\epsYt\tgb}{1\!+\!\eps\tgb}\omega_2
 \!\bigg) +
 \nonumber \\
 \!\!\!
f_{\tilde g}^{(1)}\!(x)
 \!\bigg(\!\frac{2\epsR(\epsL\omega_1\!-\!\epsY\omega_2)\tgb^2}{(1\!+\!\eps\tgb)^2} \!-\! 
\frac{\epsLR\tgb}{1\!+\!\eps\tgb}\omega_1
 \!\bigg)\!\bigg]\,,
\label{gluinoEDM}
\end{eqnarray}
where $x\!=\!m_{\tilde g}^2/m_{\tilde{q}}^2$ and $f_{\tilde g}^{(1,2,3)}(x)$
are $-2/27(-5/18)$, $2/45(7/60)$, and $-4/135(-11/180)$ for (C)EDM, respectively.
The first contribution in Eq.~(\ref{gluinoEDM}) refers to the LO contribution
proportional to the JI $J^{(d_i)}_{LR}$.  The second and third lines of
Eq.~(\ref{gluinoEDM}) contain pure BLO terms. As regards the LO term, we have
included a resummation factor, whose impact is very sizable.

The first $H^\pm$ effects to the (C)EDMs appear at the BLO~\cite{EDMH}.
In the present work, in addition to the contributions discussed in Ref.~\cite{EDMH},
we have evaluated the entire set of BLO effects that are well approximated by
\begin{eqnarray}
 \left\{\frac{d_{d_i}}{e},~d^c_{d_i}\right\}_{H^\pm}\!\!\! ~\simeq~
 \!\frac{\alpha_2}{16\pi}
 \frac{m_b}{m_{H^\pm}^2}
 \frac{m_t^2}{m_W^2}
 \frac{\epsilon_R\tgb}{(1\!+\!\eps\tgb)^2}f_{H^\pm}(z)\times
\nonumber\\
\bigg[ (1\!-\!\eps\tgb)\,\omega_2 + \epsilon_L\tgb\,\omega_1 \bigg]\,,
\label{Hedm}
\end{eqnarray}
where $z=m_t^2/m_{H^\pm}^2$ and $f_{H^\pm}(z)$ is such that
$f_{H^\pm}(1)=7/9(2/3)$.  We note that Eq.~(\ref{Hedm}) receives dominant
effects both from $J^{(d_i)}_{RR}$ and $J^{(d_i)}_{LR}$.

Charginos contribute to $J^{(d_i)}_{LL}$ already at the LO, so, the 
corresponding (C)EDMs are suppressed by $m_{d_i}$. At the BLO, a new 
effect proportional to $J^{(d_i)}_{RR}$ (and thus proportional to $m_{b}$) 
is generated by the charged-Higgsino/squark diagrams leading to
\begin{eqnarray}
 \left\{ \frac{d_{d_i}}{e},~d^c_{d_i} \right\}_{{\tilde \chi^{\pm}}}
 \!\!\!\!\!\!~\simeq~\!\!\!\!
 \frac{\alpha_2}{16\pi}
 \frac{m_b}{m_{\tilde{q}}^2}
 \frac{m_t^2}{m_W^2}\frac{A_t \mu}{m_{\tilde{q}}^2}
 \frac{\omega_2\epsilon_R \ttgb}{(1\!+\!\eps\tgb)^2}
 f_{{\tilde \chi}}\!(y)\,,
\label{charginoEDM}
\end{eqnarray}
where $\mu$ is the $\tilde{H}^\pm$ mass, $y=\mu^2/m_{\tilde{q}}^2$ and
$f_{{\tilde \chi}}(1)=-5/18(-1/6)$.
\begin{figure}[t]
\begin{center}
\includegraphics[scale=0.6]{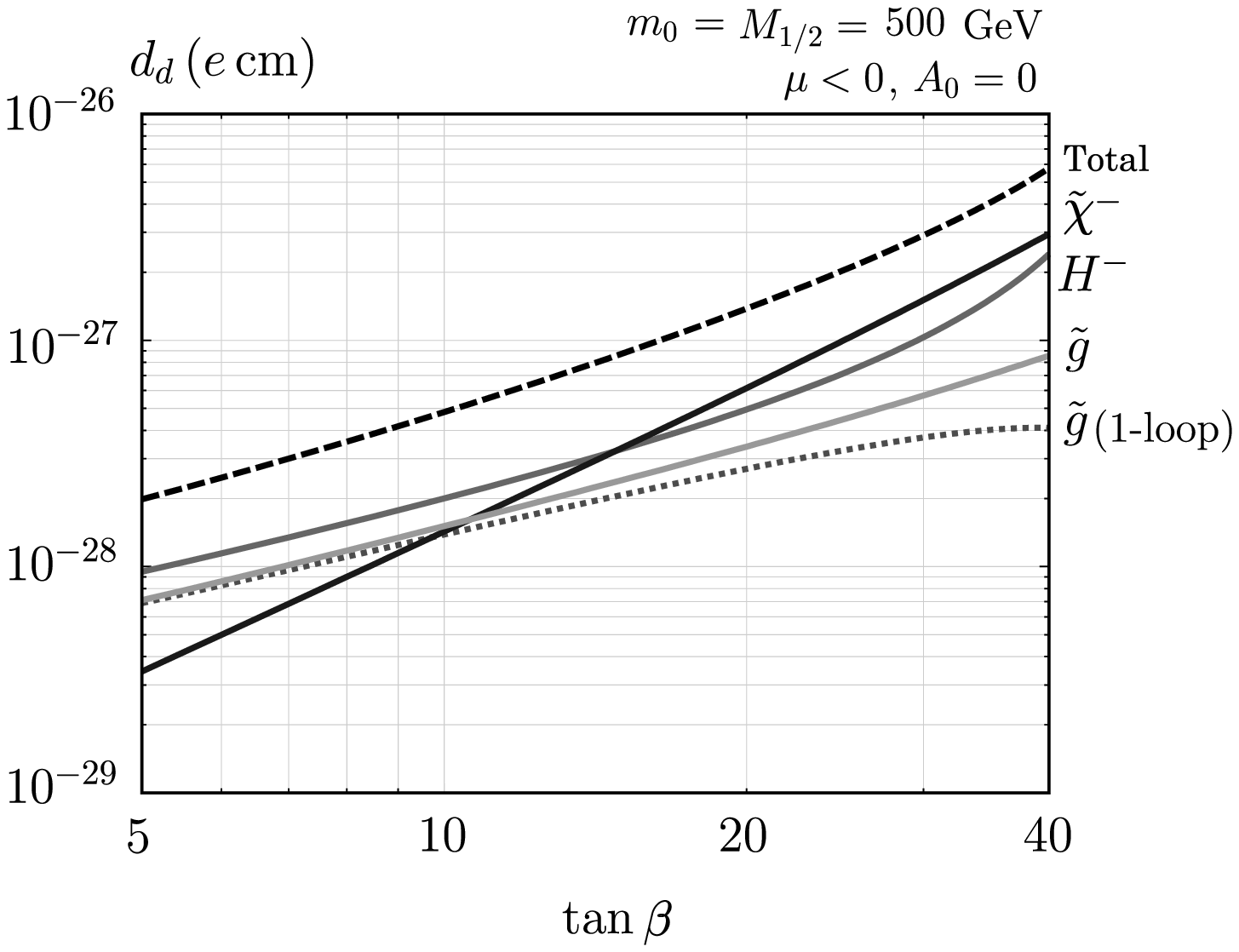}
\caption{\label{various} Contributions to the down-quark EDM assuming a CMSSM
spectrum with $m_0=M_{1/2}= 500~\rm{GeV}$, $\mu<0$, $A_0=0$, $\delta^{d}_{RR}=V$
and $\delta^{d}_{LL}=0$ at the GUT scale.
}
\end{center}
\end{figure}

The above expressions for the EDMs, {\it i.e.}, Eqs.~(\ref{gluinoEDM})-(\ref{charginoEDM}),
have been obtained by inserting effective (one-loop induced) vertices into the one-loop
expressions for the EDMs. We would like to note that such an approach accounts for all
the non-decoupling ($\tan\beta$-enhanced) contributions to the EDMs but it cannot provide
the full set of two-loop effects. The latter requires a full diagrammatic calculation,
which is outside the scope of this work.

For instance, the expression of Eq.~(\ref{Hedm}) is valid as long as
the typical supersymmetric scale $M_{\rm SUSY}$ is sufficiently larger
than the electroweak scale $m_{\rm weak}~(\sim m_W,~m_t)$ and the mass
of the charged Higgs boson $m_{H^\pm}$. Therefore, our results can be
regarded as the zeroth-order expansion in the parameters $(m_{\rm
  weak}^2,m_{H^\pm}^2)/M_{\rm SUSY}^2$ of the full
computation. However, it has been shown in Ref.~\cite{borzumati} that
this zeroth-order approximation works very well, at least in the $b\to
s\gamma$ case, even for $m_{H^\pm}\geq M_{\rm SUSY}$, provided $M_{\rm
  SUSY}$ is sufficiently heavier than $m_{\rm weak}$. This finding
should also hold in our case, considering that both $b\to s\gamma$ and
the EDMs arise from a similar dipole transition.

Moreover, in the present analysis, we have neglected potentially
relevant two-loop effects, proportional to large logarithms of the
ratio $M_{\rm SUSY}/m_{\rm weak}$, stemming from: {\it i)} the
different renormalizations of Yukawa couplings in the Higgs/Higgsino
vertices, and {\it ii)} the anomalous dimensions of the magnetic and
chromo-magnetic effective operators. These two classes of terms become
important when the scale of the supersymmetric colored particles is
significantly higher than the $W$ boson and top quark masses. 

However, we emphasize that the new two-loop effects we are dealing
with in the present work are comparable to and often larger than the
leading one-loop effects to the EDMs induced by gluino/squarks
loops. Thus, a full inclusion of all the two-loop effects to the EDMs
is not compulsory (although desirable) in a first approximation.

In order to appreciate the impact of these new contributions for the EDMs,
let us compare the size of BLO and LO effects. In Fig.~\ref{various},
we assume a CMSSM spectrum with GUT scale conditions $m_0\!=\!M_{1/2}\!=\!500~\rm{GeV}$,
$\mu<0$, $A_0\!=\!0$, $\delta^{d}_{RR}\!=\!V$ and $\delta^{d}_{LL}\!=\!0$; at the
low scale, a $\delta^{d}_{LL}\neq 0$ of order $\delta^{d}_{LL}=-|c|\,V$ 
(with $|c|\sim\mathcal{O}(0.1)$) is generated by renormalization-group (RG) effects
driven by the CKM. The above scenario finds a natural framework within SUSY GUTs with
right-handed neutrinos.

As shown in Fig.~\ref{various}, the $H^{\pm}$ and $\tilde\chi^{\pm}$
contributions are typically comparable to the $\tilde{g}$ ones.
This is possible because {\it i)} the $H^{\pm}$ and $\tilde\chi^{\pm}$
masses (entering in BLO effects) are lighter than the $\tilde{q}$ and
$\tilde{g}$ masses (entering in LO effects) in most of the MSSM parameter
space; {\it ii)} $\delta^{d}_{LL}\lsim V$ when $\delta^{d}_{LL}$ is 
radiatively-induced; {\it iii)} the mass functions for the LO $\tilde{g}$
contributions, $f_{\tilde g}^{(3)}(x)$, are more suppressed than those for
BLO $H^{\pm}$ and $\tilde\chi^{\pm}$ contributions.

Moreover, when $\tan\beta$ is large, BLO effects become more significant and
they dominate over the LO ones. In the $\tilde\chi^{\pm}$ case, this is 
explained by the explicit $\tan\beta$-dependence of
$(d_{d})_{{\tilde\chi^{\pm}}}\sim\tan^2\beta$ (see Eq.~(\ref{charginoEDM})),
to be compared with $(d_{d})_{\tilde g}\sim\tan\beta$ (see Eq.~(\ref{gluinoEDM})).
In the $H^{\pm}$ case, in spite of the same explicit $\tan\beta$-dependence of
$(d_{d})_{\tilde g}$ and $(d_{d})_{H}$, $(d_{d})_{H}>(d_{d})_{\tilde g}$ for
increasing $\tan\beta$, since $m_H$ is reduced by large RG effects driven by
$y^2_b\sim y^2_t\sim1$.

Notice that, the corner of the CMSSM parameter space where the BLO
$H^\pm$ effects are particularly enhanced compared to the LO ones,
corresponds to the{\em A-funnel} region (where $m_A\simeq 2\,m_{\rm LSP}$),
satisfying the WMAP constraints.
If we allow non-universality between the Higgs and the sfermion masses
at the GUT scale (the so-called NUHM scenarios), we can typically get
a charged Higgs that is lighter than in the CMSSM case and thus the BLO
effects become even more important.

The allowed size for hadronic EDMs is obtained by imposing the constraints 
arising from both flavor-conserving observables as $(g-2)_\mu$, $\Delta\rho$,
$m_{h^0}$ and flavor changing processes like $B\to X_s\gamma$, $B\to\tau\nu$,
$B_s\to\mu^+\mu^-$, $K_L\to\mu^+\mu^-$, $B-\overline{B}$ and $K-\overline{K}$
mixings \cite{IP}. Referring to the example of Fig.~\ref{various}, all the
above constraints are satisfied at the $99\%~ \rm{C.L.}$ for the entire
range of $\tan\beta$.

The ``flavored'' (C)EDMs have strong correlations with FCNC
observables.  As an interesting example, let us mention that,
irrespective to the particular choice for the SUSY spectrum, the
$\tilde{\chi}^\pm$ and $H^\pm$ contributions to the EDMs are closely
related to the NP contributions entering $B\to X_s\gamma$ as
\beq
(d_{d})_{\tilde{\chi}^\pm}+ (d_{d})_{H^\pm}
\simeq -e\frac{\alpha_2}{4\pi}\frac{m_b}{m^2_W}
\frac{\epsR\tgb}{1+\eps\tgb}\,\omega_2\,C_7\,,
\label{correlation}
\eeq
where $C_7 = C^{\tilde{\chi}^\pm}_{7} + C^{H^\pm}_{7}$ is defined as
${\cal B}(B\!\to\! X_s \gamma)\simeq 3.15 - 8\,C_7 - 1.9\,C_8$ \cite{Misiak}.
A detailed exploration of the intriguing correlation and interplay among FCNC
processes and flavored (C)EDMs deserves a dedicated study that goes beyond the 
scope of this Letter.

In contrast to the hadronic sector, BLO effects to the leptonic EDMs
have a rather small impact, thus, they can be neglected in the first
approximation.

In conclusion, our Letter shows that, a correct prediction for the
hadronic EDMs within SUSY theories with flavor-changing (but not
necessarily CP-violating) soft terms, necessarily requires the
inclusion of the BLO contributions presented in this work. In fact,
they do not represent just a sub-leading correction to the LO effects,
as it typically happens for flavor physics observables, but they
provide the dominant effect in large portions of the SUSY parameter
space. We also emphasize the importance of further improving the
experimental sensitivity on the hadronic EDMs as a particularly
interesting and promising probe of New Physics effects.\\

\textit{Acknowledgments:} The work of JH is supported in part by the
Grant-in-Aid for Science Research, Ministry of Education, Science and
Culture, Japan (No.~19034001 and No.~18034002). The work of MN is also
supported in part by JSPS.  The work of PP is supported in part by the
EU Contract No.~MRTN-CT-2006-035482, ``FLAVIAnet'' and by the Spanish
MEC and FEDER under grant FPA2005-01678.


\begin{thebibliography}{999}
\footnotesize{

\bibitem{pospelov}
For a review of EDMs please see,
  M.~Pospelov and A.~Ritz,
Annals Phys.\  {\bf 318}, 119 (2005) and therein references.

\bibitem{expedm}
  C.~A.~Baker {\it et al.},
  Phys.\ Rev.\ Lett.\  {\bf 97}, 131801 (2006).

\bibitem{peccei}
  R.~D.~Peccei and H.~R.~Quinn, Phys.\ Rev.\ Lett.\ {\bf 38}, 1440 (1977).

\bibitem{Jarlskog}
  C.~Jarlskog,  Phys.\ Rev.\ Lett.\  {\bf 55}, 1039 (1985).

\bibitem{GGMS}
  L.~J.~Hall et al., Nucl.\ Phys.\  B {\bf 267}, 415 (1986);
  F.~Gabbiani et al., Nucl.\ Phys.\ B{\bf 477}, 321  (1996).

\bibitem{MFV0}
L.~J.~Hall and L.~Randall, Phys.\ Rev.\ Lett.\  {\bf 65} (1990) 2939.

\bibitem{MFV}
G.~D'Ambrosio et al., Nucl.\ Phys.\ B {\bf 645}, 155 (2002).

\bibitem{EFFL}
%  K.~S.~Babu and C.~F.~Kolda, Phys.\ Rev.\ Lett.\  {\bf 84},  228  (2000);
  A.~J.~Buras et al., Nucl.\ Phys.\ B {\bf 659}, 3 (2003);
  A.~Dedes and A.~Pilaftsis, Phys.\ Rev.\ D {\bf 67},  015012 (2003);
  J.~Foster et al., JHEP \ {\bf 0508},  094 (2005).


\bibitem{HRS}
L.~J.~Hall, R.Rattazzi and U.~Sarid, Phys.\ Rev.\ D {\bf 50}, 7048 (1994).


\bibitem{EDMH}
  J.~Hisano and M.~Nagai and P.~Paradisi,
  Phys.\ Lett.\ B {\bf 581},  224 (2006).


\bibitem{borzumati}
  F.~Borzumati {\it et al.}, Phys.\ Rev.\ D {\bf 69},  055005 (2004).


\bibitem{IP}
  For a recent analysis please see, G.~Isidori and P.~Paradisi,
  Phys.\ Lett.\ B {\bf 639},  499 (2006).

\bibitem{Misiak}
  M.~Misiak {\it et al.}, Phys.\ Rev.\ Lett.\  {\bf 98},  022002 (2007);
  M.~Misiak and M.~Steinhauser, Nucl.\ Phys.\  B {\bf 764},  62 (2007);
  M.~Misiak and M.~Steinhauser, private communication.

}
\end{thebibliography}
\end{document}